\documentclass[%
aip,
sd,%
amsmath,amssymb,
reprint,%
]{revtex4-1}

\usepackage{graphicx}% Include figure files
\usepackage{dcolumn}% Align table columns on decimal point
\usepackage{bm}% bold math
\usepackage{pgf}
\usepackage{ragged2e}
\usepackage{tikz}
\usepackage{verbatim}
\usepackage{tikz} 
\usetikzlibrary{arrows,shapes,snakes,automata,backgrounds,petri}
\usepackage[latin1]{inputenc}
\usepackage{mathrsfs}  
\usepackage[noend]{algpseudocode} 
\usepackage{hyperref}
\usepackage{mathrsfs, comment, dsfont, esint, mathtools, subfigure, url,}
\usepackage{amsthm}
\usepackage{algorithm}
\usepackage{pgfplots}
\pgfplotsset{compat=1.8}
\usepgfplotslibrary{statistics}

\newcommand{\spec}[1]{\langle #1 \rangle}

\newcommand{\tr}[1]{\mathrm{tr}\left( #1 \right)}

\newcommand{\vect}[1]{ \mathbf{ #1} }

\def\comm#1{\texttt{\color{gray} #1}}  
\newcommand{\floor}[1]{ \left \lfloor{#1}\right \rfloor}

\begin{document}
    
    \preprint{AIP/123-QED}
    
    \title{QUBO-based density matrix electronic structure method}% Force line breaks with \\
    % \thanks{Footnote to title of article.}
    \author{Christian F. A. Negre, Alejandro Lopez-Bezanilla, Yu Zhang, Prosper D. Akrobotu, Susan M. Mniszewski, Sergei Tretiak, and Pavel A. Dub}
    \begin{abstract}
        Density matrix electronic structure theory is used in many quantum chemistry methods to ``alleviate" the computational cost that arises from directly using wave functions. Although density matrix based methods are computationally more efficient than wave functions based methods, yet significant computational effort is involved. Since the Schr\"odinger equation needs to be solved as an eigenvalue problem, the time-to-solution scales cubically with the system size, and is solved as many times in order to reach charge or field self-consistency. We hereby propose and study a method to compute the density matrix by using a quadratic unconstrained binary optimization (QUBO) solver. This method could be useful to solve the problem with quantum computers, and more specifically, quantum annealers.
        The method hereby proposed is based on a direct construction of the density matrix using a QUBO eigensolver. We explore the main parameters of the algorithm focusing on precision and efficiency. We show that,
        while direct construction of the density matrix using a QUBO formulation is possible, the efficiency and precision have 
        room for improvement. Moreover, calculations performing Quantum Annealing with  the D-Wave's new Advantage quantum processing units is compared with classical Simulated annealing, further highlighting some problems of the proposed method. 
        We also show some alternative methods that could lead to a better performance of the density matrix construction.
    \end{abstract}
    
    \maketitle
    
    \section{Introduction}
\begin{figure}

\begin{tikzpicture}[node distance=1.3cm,>=stealth',bend angle=45,auto]

  \tikzstyle{place}=[rectangle,thick,draw=blue!75,fill=blue!20,minimum size=6mm]
  \tikzstyle{red place}=[place,draw=red!75,fill=red!20]
  \tikzstyle{transition}=[rectangle,thick,draw=black!75,
  			  fill=black!20,minimum size=4mm]

  \tikzstyle{every label}=[red]

  \begin{scope}
    % First net
    \node [place] (h0)                                    {$H = H^0 + V$};
    \node [] (c1) [below of=h0, draw=white]                      {};
    \node [place] (s)  [align=center,below of=c1,text width=3.0cm,centered] {if $|\rho-\rho_{old}| < tol$,   STOP};

    \node [transition] (e1) [left of=c1,xshift=-0.5cm] {$\rho = C f_F(\epsilon) C^{\dagger}$}
      edge [pre,bend left]                  (h0)
      edge [post,bend right]                (s);

    \node [transition] (l1) [right of=c1,xshift=0.5cm] {$V = f_C(\rho)$}
      edge [pre,bend left]                  (s)
      edge [post,bend right] node[swap] {} (h0);

  \end{scope}

  \begin{scope}[xshift=10cm]
    % Second net
    \node [place]     (c1') [right of=c1, xshift=4cm]   {$\vect{x}^t Q \vect{x}$}; 
  \end{scope}

  \draw [-to,thick,snake=snake,segment amplitude=.4mm,
         segment length=2mm,line after snake=1mm]
    (l1) -- (c1')
    node [above=1mm,midway,text width=1cm,text centered, xshift=0.0cm]
      {\textcolor{red}{Paradigm shift}};

\end{tikzpicture}
\caption{Scheme showing the paradigm shift hereby proposed. On the left we represent the traditional self-consistent charge (SCC) approach with steps that include the correction of the Hamiltonian elements $H$; the construction of the density matrix $\rho$ from a Fermi function $f_F$ of the Hamiltonian; and the calculation of the effective potential $V$ as a Coulombic function $f_C$ of $\rho$ to be used to reconstruct the Hamiltonian in the subsequent SCC step. On the right, the alternative QUBO idea is proposed.}
\label{scheme}
\end{figure}
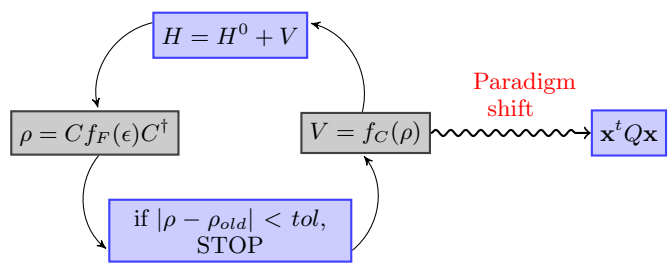

    The correct choice of the model or method used to represent the electronic structure of a chemical system is crucial to get a good description of any observable (forces, energies, etc.) of such a system. The quality of a model can be estimated as the ratio between its predictive power
    and its complexity (\textit{Quality} = \textit{Predictive Power}/\textit{Complexity}), the predictive power being related to the amount of information that can be extracted from a calculation, and the complexity mainly referring to how the time-to-solution scales with the system size \cite{Finnis2004}. 
    Although this definition is vague, it gives us the general understanding that, in order to optimize a method (increase its quality) it is necessary to work on approaches that could decrease the complexity, and, at the same time, avoid losing predictive power; In many cases the predictive power translates into accuracy. 
    Reducing the complexity is the typical approach that has been taken given that computational resources have always been limited due to the ever increasing need of addressing larger systems. However, with the arrival of quantum computers (an alternative computational paradigm) and more specifically, quantum annealers (QAs) \cite{dwave}, it is possible that more complex formulations (in terms of traditional computation) of a given problem end up having shorter or even instantaneous time-to-solution without accuracy resignation. This has been the case of graph problems including graph partitioning and community detection solved with QAs that was made possible due to a quadratic unconstrained binary optimization (QUBO) reformulation of the problem \cite{Ushijima-Mwesigwa2017-li,Negre2019-yh}. 
    For the case of quantum chemistry methods, the long term vision to reach is the one in which all the computational burden of the eigenvalue problem and self-consistent process could be shifted towards an ``extremely complex,'' yet easily solvable QUBO problem (See Figure \ref{scheme}). 
    In this paper we take an initial step towards the aforementioned vision by exploring how feasible it is to construct the density matrix (DM) using linear algebra QUBO algorithms. Previous work on QAs used in quantum chemistry has been recently applied to determine the ground and firsts excited states on the Full Configuration Interaction (CI) method by diagonalizing the Full CI (FCI) matrix \cite{Teplukhin2020-rm}; the downfolding of the FCI matrix to reduce the size of the effective matrix \cite{Mniszewski2021-ve}; and Time Dependent Density Functional Theory (DFT)  to compute excitations \cite{Teplukhin2021-la}. 
    
    The density matrix (DM) formalism is ubiquitous to several methods in computational chemistry that compute the 
    electronic structure of a molecular system. 
    Among these methods, the Khon-Sham method within DFT is arguably the most frequently used of the quantum chemistry methods,
    preferred over post Hartree-Fock methods, for its performance in terms of predictive power relative to its computational cost \cite{VazquezMayagoitia2014-dg,Fiolhais2003}. The foundations of this theory will be briefly stated below.
    
    The many-body wave function of interacting electrons in a chemical system resulting in the solution of the Schr\"odinger equation depends on 3$\times N_e$ spatial coordinates, where $N_e$ is the number of electrons in the system. DFT can be used to overcome this complexity, 
    allowing for the calculation of observables just by knowing the electronic density of the system. 
    This theory is based on using the electronic density $n(r)$  ($r$ being the electron coordinates) as a variable of the total energy functional $E[n(r)]$ of the system, and the variational principle to minimize the latter. 
    The theorems of P. Hohenberg and W. Kohn \cite{Kohn1964} ensure that the energy, wave function, and all other electronic properties are uniquely determined by the electron density $n (r)$. Moreover, for any external potential 
    $v_{ext}(r)$ generated by the nuclear configuration of a molecular system, there is a bijective relation with a unique density function $n(r)$. To solve for $n(r)$ the Kohn-Sham method relies on auxiliary wave functions $\psi_i(r)$ that are used to transform the minimization of the energy functional into the following eigenvalue problem:
    \begin{align}
        \left(-\frac{\nabla^2}{2} + v_{eff}\right) \psi_{i}(r) = \epsilon_i\psi_i(r) \\
        \hat{H} \psi_{i}(r) = \epsilon_i\psi_i(r)
        \label{khonshameigen}
    \end{align}
    where: 
    \begin{equation}
        v_{eff}=v_{ext}(r) + \int \frac{n(r')}{|r-r'|}dr' + \frac{\delta E_{xc}[n(r)]}{\delta n(r)}
        \label{veff}
    \end{equation}
    Here, $v_{ext}(r)$ is the external potential given by the nuclei; It is the part of the energy functional that characterizes the chemical system in question. 
    $E_{xc}$ is the functional that takes into account the energy of exchange-correlation. It
    is written as the sum of two components: The kinetic energy correlation component, which comes from correcting the error by considering the kinetic energy of non-interacting electrons, and the inter-electronic interaction
    containing the non-kinetic correlation and exchange. The exact form of this functional is not known, but many approximations exists nowadays with varying accuracies and computational costs \cite{Sousa2007-zq}. In this work we will use the most basic one where the functional is
    represented by a local density
    approximation (LDA) of the form: $E_{xc}[n(r)] = \int \epsilon_{xc}(n)n(r)d^3r$, where $\epsilon_{xc}(n)$ is the correlation and exchange energy per electron for a uniform gas of density $n$.
    Equation \ref{khonshameigen} is solved iterativelly using equation \ref{veff} until reaching self-consistency between the density $n(r)$ and the effective potential $v_{eff}(r)$.

    When written using a basis set of functions $\{\phi_i\}_{i \in 1,..,N}$ with $S_{ij} = \int_r \phi^{*}_i(r) \phi_j(r) dr$  and $H_{ij} = \int_r \phi^{*}_i(r) \hat{H} \phi_j(r) dr$ equation \ref{khonshameigen} turns into the following generalized eigenvalue problem:
    \begin{equation}
        H \vect{c} = \epsilon_i S\vect{c}
        \label{singlev}
    \end{equation}
    where, $c_{k}$ is the k-th component of the expansion of the wavefunction $\psi_i(r)$ in the chosen basis set. In matrix form equation \ref{singlev} becomes: 
    \begin{equation}
        H C = \epsilon S C 
    \end{equation}
    which is a generalized matrix eigenvalue problem, where $\epsilon$ is a diagonal matrix with entries $\epsilon_{ij} = \delta_{ij} \epsilon_i$ containing all the eigenvalues of matrix $H$. Finally, $C$ is a matrix containing all the expansion coefficients for each of the $N$ eigenfunctions $\psi_i(r)$. 
    
    Given the expansion coefficients $C$, the DM which is our object of interest here, is constructed as: 
    \begin{equation}
        \rho=Cf(\epsilon)C^{t}
    \end{equation}
    where we have used the Fermi distribution function $f_n$  for a system in thermal equilibrium defined as:
    \begin{equation}
        f_n(\epsilon)= \frac{2}{1+\exp(\frac{\epsilon-\epsilon_F}{k_bT})}
        \label{Fermi}
    \end{equation}
    where $ k_b $ is the Boltzman constant, $ \epsilon_F $ is the Fermi level (or electrochemical potential) and $ T $ is the system's temperature.
    For practical purposes, when temperature is low, $ f_n(\epsilon) = 2\Theta(\epsilon - \epsilon_F)$, leading to: 
    \begin{equation}
        \rho = \sum_i^{occ} 2 \vect{c}_i \vect{c}_i^{T}
        \label{Fermi}
    \end{equation}
    The DM can be hence directly constructed by summing the self outer products of each eigenvector (in order of increasing energy $\epsilon_i$) up to the number of occupied states ($occ = N_e/2$).
    Given the DM, many properties can be computed since the expectation value of any operator $ \hat{A} $ can be calculated
    as $\spec{A} = tr(\rho A)$. For instance, the trace of the DM gives the total number of electrons of the system, and the trace of $\rho H$  provides the total electronic energy also known as \textit{band} energy of the system \cite{Szabo1996}.
    Another important property is that we can easily compute $n(r)$ as follows:
    \[
    \begin{split}
        n(r) =     
        \sum_{k,l}
        \phi_k(r) \rho_{kl} \phi_l(r) 
    \end{split}
    \]
    which can be reused in equation \ref{veff} to recompute $\hat{H}$.
    Other methods such as Hartree-Fock, Tight-Binding, and Semiempirical  are also based on constructing the single-particle DM \cite{Elstner1998,RMcWeeny56,RMcWeeny60,Finnis2004}.
    From what was explained above, matrices  $\epsilon$ and $C$ arise from diagonalizing the matrix representation of $ \hat{H} $ in the aforementioned basis set.
    This diagonalization step in all the density matrix based methods is typically the bottleneck of the whole calculation, and scales as $\mathcal{O}(N^3)$, where $N$ is the number of elements in the basis set (which scales linearly with the system size). 
    Over the years, computational chemists have dedicated lots of effort to develop algorithms that could solve the density matrix with $\mathcal{O}(N)$ scaling \cite{sgoedecker03, bowler1999-yu,anders_2002}; leading to the so called ``order n methods.'' Some of them have also adapted algorithms to the ever evolving classical computer 
    architectures \cite{smniszewski15, Finkelstein2021-sc}.
    The problem with all these order n approaches is that they require a trade-off between accuracy and efficiency. The faster we want the method to be, the lower will be the accuracy of our results. In this paper we analyze the tractibility of quantum annealing as an alternative computational paradigm to get accurate results at a lower computational cost for DM based quantum chemistry methods.

    \section{Methods}

    In this section we will describe the methods used, and in particular we will focus on our recently developed high precision QUBO eigensolver \cite{Krakoff2021-vf}. 
    
    \subsection{Quantum annealers}
    
    Quantum Annealers (QAs) are specific adiabatic quantum computing devices particularly suitable to solve combinatorial optimization problems \cite{dwave_welcome}. In these devices the exploration of the combinatorial space is given by the delocalization and degree of superposition of the initial quantum state that decays to the state of minimum energy during a process called quantum annealing.
    The D-Wave QA is an architecture that uses this quantum annealing technique to solve for the ground state of an Ising Model Hamiltonian that is derived from a combinatorial optimization problem. In other words, any problem to be solved
    with the D-Wave QA will need to be reformulated to be solved as an Ising Model Hamiltonian whose energy is evaluated as follows:
    \begin{equation}
        \label{eq:isingenergy}
        E(\vect{s}) = \sum_{i}h_{i}s_{i} + \sum_{i<j}J_{i,j}s_{i}s_{j}
    \end{equation}
    for which $s_{i}\in \{-1,1\}$ are magnetic spin variables; $h_{i}$ are local programmable fields; and 
    $J_{ij}$ are programmable field coupling interactions. An Ising model Hamiltonian can be easily transformed into its boolean equivalent using the following transformation: $\vect{s} = 2\vect{q} - \vect{1}$, where  $\vect{1} = (1,\ldots,1)$. The boolean equivalent 
    problem that is obtained is referred to as a QUBO problem and it is expressed as follows:
    \begin{equation}
        E(\vect{q}) = \vect{q}^{T}Q\vect{q} = \sum_{i}q_{i}Q_{ii} + \sum_{i\neq j}q_{i}q_{j}Q_{ij}.  
    \end{equation}
    with $q_{i}\in \{0,1\}$.
    
    \subsection{High-precision QUBO-based eigensolver}
    
    The solution of the extremal eigenvectors with high accuracy (eigenvector corresponding to minimal or maximal eigenvalue) was recently possible thanks to a new QUBO formulation of the problem. 
    \cite{Krakoff2021-vf}. This method builds on top of the quantum annealer eigensolver (QAE) work of Teplukhin et. al. \cite{Teplukhin2020-rm}.
    The main difference between this high-precision QUBO eigensolver method and the previous QAE, is that it solves the problem as a fixed point method instead of using binary search \cite{Hoffman2018-vo}. The new method also involves a steepest descent search together with a systematic adjustment of the length of the precision vector to guarantee high accuracy convergence.
    The precision vector defined as follows is used to convert numbers from binary to real representation: 
    \[\vect{v} = (-1,2^{-1},2^{-2},...,2^{-m})   \]
    where $m$ is the number of bits used in the representation. In order to transform a number from  binary $q$ to real $x$ 
    representation we need to perform $x = \vect{v}^{T}\vect{q}$, where $\vect{q}$ is the vectorized form of $q$.  
    Note that with this representation  we cover the $[-1, 1)$ segment where the binary representation $q$ of $x$ will satisfy:
    $ -1 \equiv (1,0,0,0,...,0) \le  q <  (0,1,1,1,...,1) \equiv 1  $. 
    A real field multiplication of $x \times y$ has the binary equivalent $q_x^{T}  (\vect{v}\otimes \vect{v}^{T}) q_y$. A particular example of this multiplication is given in the Appendix.
    
    With the help of this precision vector and the previously introduced multiplication, any quadratic equation $\vect{x}^{T} B \vect{x}$ for which we need to find an optimal $\vect{x}$ with $B \in  \mathbb{R} ^{N\times N}$, will have binary equivalent   
    $\vect{q}^{T} (B  \otimes (\vect{v} \otimes \vect{v}^{T})) \vect{q} $, where $(B  \otimes (\vect{v} \otimes \vect{v}^{T})) \in \mathcal{B}^{mN\times mN}$, being $\mathcal{B}^{mN\times mN}$ the ${mN\times mN}$ binary field vector space.
    Any linear term $\vect{r}^{T} \vect{x} $ with  $\vect{r} \in  \mathbb{R} ^{N}$ for which we need to find an optimal $\vect{x}$, will have a binary equivalent  $\vect{q}^{T} \mathrm{Diag}(\vect{r} I \otimes \vect{v}))\vect{q} $, where $I$ is the $N \times N$ identity matrix, and $\mathrm{Diag}()$ is an operator that, given $\vect{a}$ in $\mathcal{B}^{mN} $ builds an operator $A$ in $\mathcal{B}^{mN\times mN} $ with $A_{ij}= \delta_{ij}a_i$. 
    
    Given the Hamiltonian ($H$) and overlap ($S$) matrices, every eigenpair $(\lambda,\vect{x})$ will satisfy: 
    $ (H\vect{x} - \lambda S\vect{x}) = 0$, and, in particular, the lowest eigenpair ($\lambda_{min}$, $\vect{x}_{min}$) will be an absolute minimum of: 
    \begin{equation}
        (H\vect{x} - \lambda S\vect{x})^{T} (H\vect{x} - \lambda S\vect{x})
    \end{equation}
    Since $H - \lambda S$ is positive semidefinite, 
    we can instead minimize: 
    \begin{equation}
        \vect{x}^{T} (H  - \lambda S)  \vect{x}  \quad \equiv  \quad \vect{q}^{T} (H  - \lambda S)\otimes (\vect{v} \otimes \vect{v}^{T})  \vect{q}
        \label{eq1}
    \end{equation}
    Since $\lambda_{min}$ is not known, equation \ref{eq1} needs to be solved iterativelly as a fixed point. A good initial value for $\lambda$ happens to be the average eigenvalue of $H$ computed as $\tr{H}/N$ \cite{Krakoff2021-vf}. The method converges once the changes in $\lambda_{min}$ are below a certain preset tolerance. 
    
    In order to achieve a higher precision in the calculation of both $\vect{x}_{min}$ and $\lambda_{min}$, we have introduced a steepest descent phase also solved as a QUBO problem.  Setting the  previously converged  $\vect{x}_{min}$ as an initial guess $\vect{x}^0_{min}$, we Taylor expand $f(\vect{x}) = \vect{x}^{T} (H- \lambda S)\vect{x}$ around $\vect{x}^0_{min}$ to obtain:
    \begin{align*}
        f(\vect{x}) & = f(\vect{x}^0_{min}) + \nabla f \cdot \delta + \delta^{T} \frac{\mathcal{H}(f)}{2} \delta + o(||\delta||^3) \\
        \delta & := \vect{x} - \vect{x}^0_{min}
    \end{align*}
    where $ \nabla f$ and  $\mathcal{H}(f)$ are the gradient and Hessian of $f$ respectivelly.
    For the case of $f(\vect{x})$, the gradient and Hessian becomes  $\nabla f = 2(H-\lambda S)(\vect{x}^{0}_{min})^{T}$, and  $\mathcal{H}(f) = 2(H - \lambda S)$ respectivelly. Hence, we can compute a high-precision approximation to the lowest eigenpair as: 
    \begin{align*}
        \vect{x} & = \vect{x}^0_{min} + \underset{\vect{\delta}}{\text{argmin}}\left( \nabla f \cdot\vect{\delta} +  \vect{\delta}\cdot \mathcal{H} \vect{\delta}\right)
    \end{align*}
    In this case, $\vect{\delta}$ converges to a fixed point, and, once $\vect{\delta}$ does not change anymore, the length of the precision vector is shortened to 10\% of its previous value. This procedure guarantees convergence to any arbitrary desired precision. 
    
    For every ``next'' eigenpair we want to compute, the previous eigenvalue $\vect{x}_{min} \equiv \vect{x}_{i-1}$ of $H$ needs to be ``pushed out'' of the eigenspectrum  by using the following transformation: 
    \begin{equation}
        H = H -  w S \vect{x}_{i-1} \vect{x}^{T}_{i-1}S  
    \end{equation}
    where $w$ is an estimation for the ``eigenspectrum width.'' A good estimation of $w$ can be calculated as $w = h - l$ where $h$ and $l$ are the highest and lowest Gershgorin bounds of $H$ respectively. 
    
    Using the previously explained procedure to compute eigenvectors, the density matrix is constructed as explained before; using a straightforward summation of eigenvector outer products up to the number of occupied states. This is: 
    \begin{equation}
        \rho  = \sum_k^{nocc} \vect{x}_k \vect{x}^{T}_k  
    \end{equation}
    
    A python like pseudocode for the full algorithm is detailed in Pseudocode 1. The inputs are the Hamiltonian $H$, Overlap $S$, the occupation $nocc$, and the target precision $prec_T$. The output is the DM $\rho$.
    
    \begin{algorithm}[H]
        \floatname{algorithm}{\scriptsize Pseudocode}
        \algrenewcommand\algorithmicfunction{\textbf{def}}
        \algrenewcommand\algorithmicend{.}
        \begin{algorithmic}
            \label{rhocode}
            \parskip 0.05cm
            {\fontsize{0.2cm}{0.2em}\selectfont 
                \Function{get\_rho}{$H,S,nocc,prec_T$}: 
                \State \comm{ \#Eigenspectrum width estimation}
                \State $w$ = \textbf{get\_Gershgorin\_width}($H$)  
                \State \textbf{for} i \textbf{in range}(nocc): \comm{\#Loop over occ states}
                \State $\lambda$ = $\tr{H}/N$  \comm{ \#Initial estimate of $\lambda$} 
                \State \comm{\#Initial phase}
                \State \qquad \textbf{while} ($err > tol$): 
%                 \State \qquad \qquad \qquad  $Q =  (H - \lambda S)$  %\otimes  M $ \comm{\#Construc QUBO}
                \State \qquad \qquad \qquad $\vect{x} = $\textbf{solve}($\vect{x}^{T}(H - \lambda S)\vect{x}$) \comm{\#Send to Annealer}  
                \State \qquad \qquad \qquad  $ \vect{x} \leftarrow \vect{x}/(\vect{x}^{T} S \vect{x})$ \comm{\#Normalize}
                \State \qquad \qquad \qquad  $\lambda =  \vect{x}^{T} H \vect{x}$ \comm{\#Compute new eigenvalue}
                \State \qquad \qquad \qquad  $err =  \vect{x}^{T} (H -  \lambda S)\vect{x}$ \comm{\#Compute error}                
                \State \comm{\#Descent phase}
                \State \qquad $prec = 0.1$
                \State \qquad \textbf{while} ($prec > prec_T$): 
%                 \State \qquad \qquad \qquad $ Q = \nabla f \delta +  \delta \mathcal{H} \delta$
                \State \qquad \qquad \qquad $\vect{\delta} = $\textbf{solve}($\nabla f \delta +  \delta \mathcal{H} \delta$) \comm{\#Send to Annealer}
                \State \qquad \qquad \qquad $\vect{x}  \leftarrow \vect{x} + \vect{\delta} $
                \State \qquad \qquad \qquad  $ \vect{x} \leftarrow \vect{x}/(\vect{x}^{T} S \vect{x})$ \comm{\#Normalize}
                \State \qquad \qquad \qquad  $\lambda =  \vect{x}^{T} H \vect{x}$ \comm{\#Compute new eigenvalue}
                \State \qquad \qquad \qquad  $err =  \vect{x}^{T} (H -  \lambda S)\vect{x}$ \comm{\#Compute error}   
                \State \qquad \qquad \qquad     \textbf{if} $(err > err_{old}): $
                \State \qquad \qquad \qquad  \qquad $\vect{v} \leftarrow 0.1 \vect{v}$
                \State \qquad \qquad \qquad  \qquad $prec = 0.1 prec$
                \State \qquad \qquad \qquad    $err_{old}  = err$
                
                \State \qquad $H \leftarrow H - w  S \vect{x}^{T} \vect{x} S$ \comm{\#Shift eigenvalue up}
                \State \qquad $\rho \leftarrow \rho + 2  \vect{x}^{T} \vect{x}$ \comm{\#Construct partial DM}
                \State    \textbf{return} $\rho$                                          
                \EndFunction
            }       
            \caption{\scriptsize Full DM construction algorithm using high-precision QUBO eigensolver. Description of the variables follows. $H$: Input Hamiltonian; $prec_T$: Target precision; $nooc$: Number of occupied states; $\lambda$: Minimal eigenvalue being computed; $N$: Size of $H$; $err$: Convergence error; $tol$: Tolerance for the initial phase; $S$: Overlap matrix; $\vect{x}$: Minimal eigenvector being computed; $prec$: Convergence precision; $\nabla f$: Gradient of $f$; $\mathcal{H}$: Hessian matrix; $\rho$: Output DM}
        \end{algorithmic}
    \end{algorithm}
    Pseudocode 1 implements a straightforward way of computing the density matrix using the high-precision QUBO eigensolver. 
    From this pseudocode we can immediately identify some important properties. We see that the matrix operations in the inner loops are $\mathcal{O}(N^2)$
    unless they can be solved using sparse matrix formats to get linear scaling. 
    The other important point is that the two inner loops (initial and descent phases) have to be solved $nocc$ times, where $nocc$ is the number of occupied states that depends on the number of electrons which scales linearly with the number of orbitals $N$. From this, we can deduce that even if the solution of the quadratic problem takes no time, the overall scaling of the algorithm gives back the ``undesired'' $\mathcal{O}(N^3)$ scaling. Provided the matrix operations in the inner loop could be solved with $\mathcal{O}(N)$ scaling methods, and the quadratic problem takes no time, we will have at best an $\mathcal{O}(N^2)$ scaling. 
    Even though this seems to be an undesired result, it is a first direct approach to solve the DM using a QUBO eigensolver and deserves some further study.

    \subsection{DFT calculations}
    
         We used the DFT-based code SIESTA  \cite{SIESTA,SIESTA2} to obtain all the molecules' Hamiltonians in a tight-binding like representation. SIESTA is a first-principles electronic structure code for molecules and solids which represents wavefunctions as a combination of atomic-like orbitals\cite{Sankey}. The Kohn-Sham equations are solved self-consistently in a linear combination of atomic orbitals basis set and they are restricted to a fix-size cell condition. This framework allows us to obtain a description of the operator associated with the system energy employing a reduced number of orbitals in the basis set with complete multiple-zeta and polarized bases, depending on the required accuracy. In this study, all first-principles calculations were performed with a minimal single-$\zeta$ basis for each atom. 
     Description of the interaction between atoms was conducted within the local density approximation approach\cite{PhysRev.140.A1133} for the exchange-correlation functional. The integration over the Brillouin zone (BZ) was performed using a Monkhorst sampling in $\Gamma$ point. The radial extension of the orbitals had a finite range with a kinetic energy cutoff of 50 meV. A separation of 20 \AA\ in a cubic simulation box prevents virtual periodic molecules from interacting. We used different numbers of molecules to get the Hamiltonians ($H$) and overlap ($S$) matrices needed to test our algorithm. We also computed $H$ and $S$ for benzene in order to test for resiliency regarding degeneracy of eigenstates.

    \section{Results and discussion}
\begin{figure}[ht]
        \centering
        \includegraphics[width=7cm]{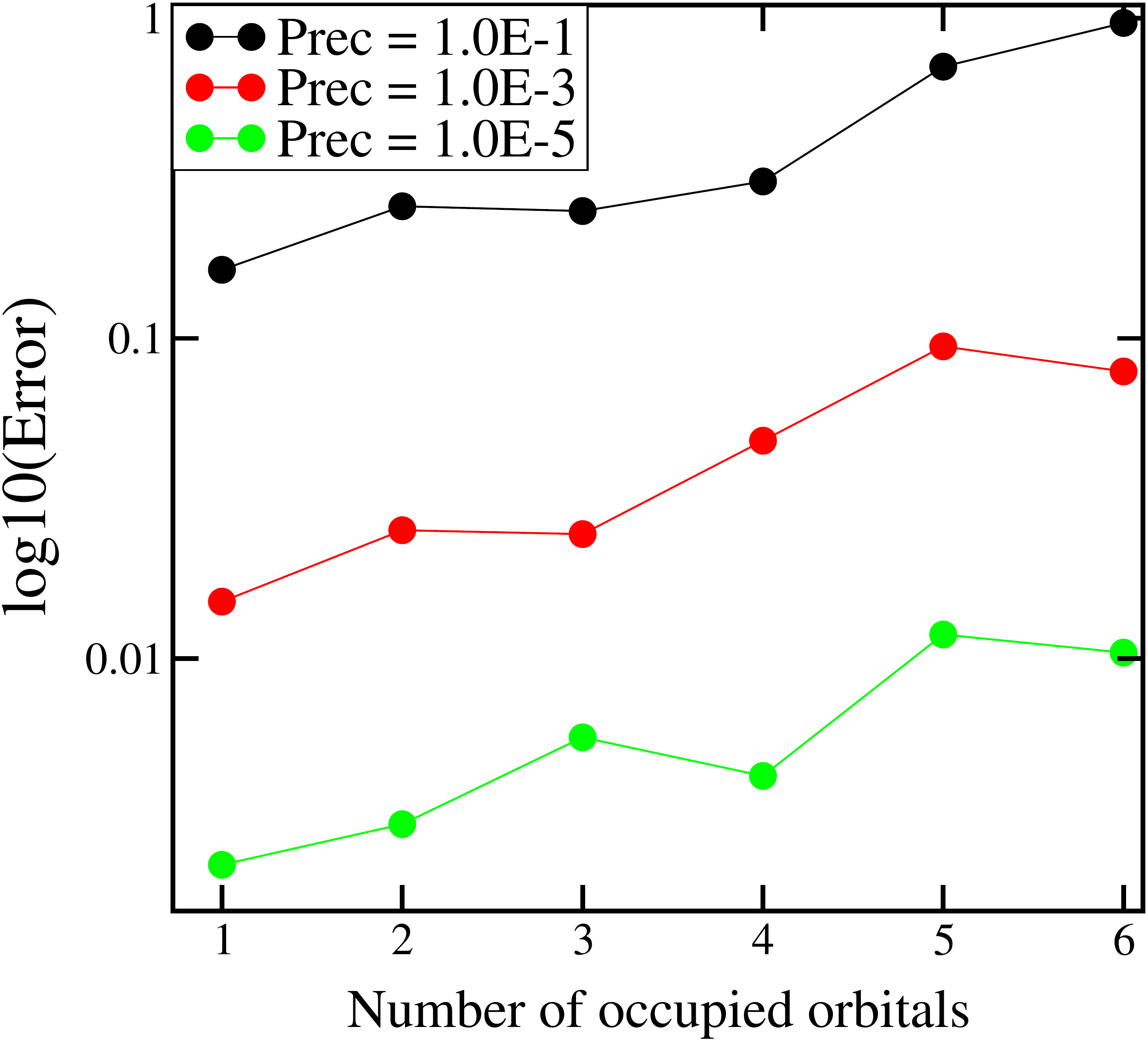}
        \caption{Error as a function of the number of occupied states for a single water molecule with  $N=6$. Results for three different precision values are shown.}
        \label{err_vs_nocc}
    \end{figure}

    We first studied the propagation of the errors with the number of occupied states. 
    The error in the calculated DM is computed as follows:
    \begin{equation}
        error =  || \rho - \rho^{ex} ||_1 = \sum_{ij} | \rho_{ij} - \rho^{ex}_{ij} |
    \end{equation}
    where $\rho^{ex}$ is the DM obtained using regular diagonalization in double precision which we hereby set as our standard.
    In order to reach chemical accuracy, the error needs to satisfy $error  \le  1.0[kCal/mol]/ \max_{lm} ( H_{lm} )$ as explained in the Appendix.
    Results showing scaling of error and iteration with system size, occupation, and degeneracy were performed using classical SA algorithm as implemented in the dwave-neal code for the Ocean API \cite{dwave-neal}. A comparison with QA running on the D-Wave Advantage 4.1 system is offered at the end of this section. 
    From the Pseudocode 1, we can deduce that the error committed on the 
    calculations of the very first computed eigenpairs is susceptible to be propagated towards the higher eigenpairs, hence, making the calculations of DM with higher occupations 
    less accurate. 
    Figure \ref{err_vs_nocc} shows the error as a function of the occupation number for a single water molecule with six total atomic orbitals. Three different precision values were used to show that this problem happens regardless of the precision used. In general, we notice that the higher we set the precision for the computation of the eigenvectors, the less error is committed in the construction of the DM across different values of occupations. The larger the number of occupied orbitals, the larger the error that is committed in the computation of the DM.  
    
\begin{figure}[ht]
        \centering
        \includegraphics[width=7cm]{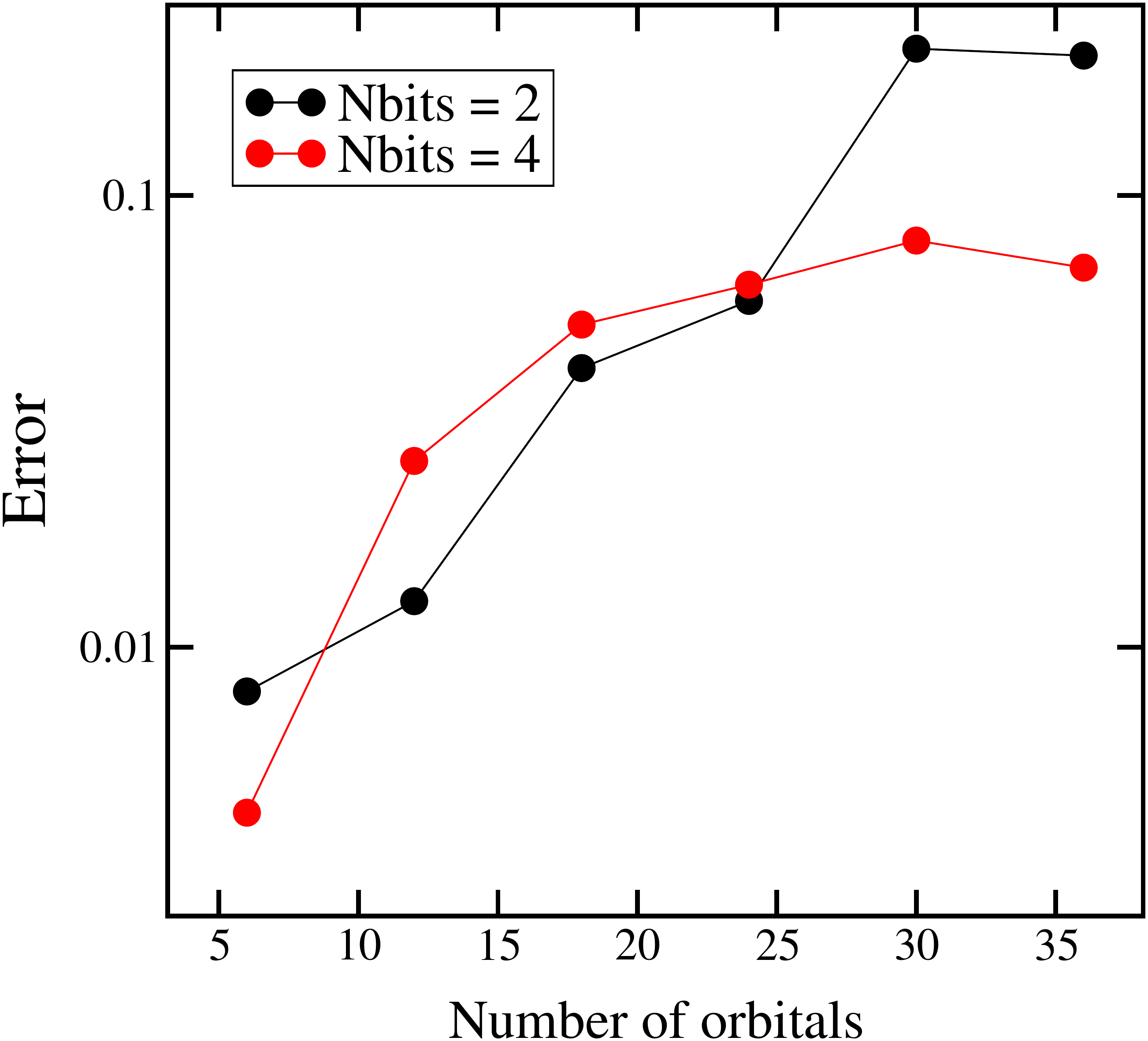}
        \caption{Error as a function of system size (number of orbitals). Calculations using two and four bits of precision are shown.}
        \label{err_vs_size}   
    \end{figure}

    The problem of error propagation worsens with system size. As it is shown in Figure \ref{err_vs_size}, 
    the error gets larger when the total number of orbitals increases, regardless of the precision and number of bits that are used. We notice that, the error increases with the system size which is a sign that a higher precision will be required for larger systems to reach chemical accuracy.   
    
\begin{figure}[ht]    
        \centering
        \includegraphics[width=7cm]{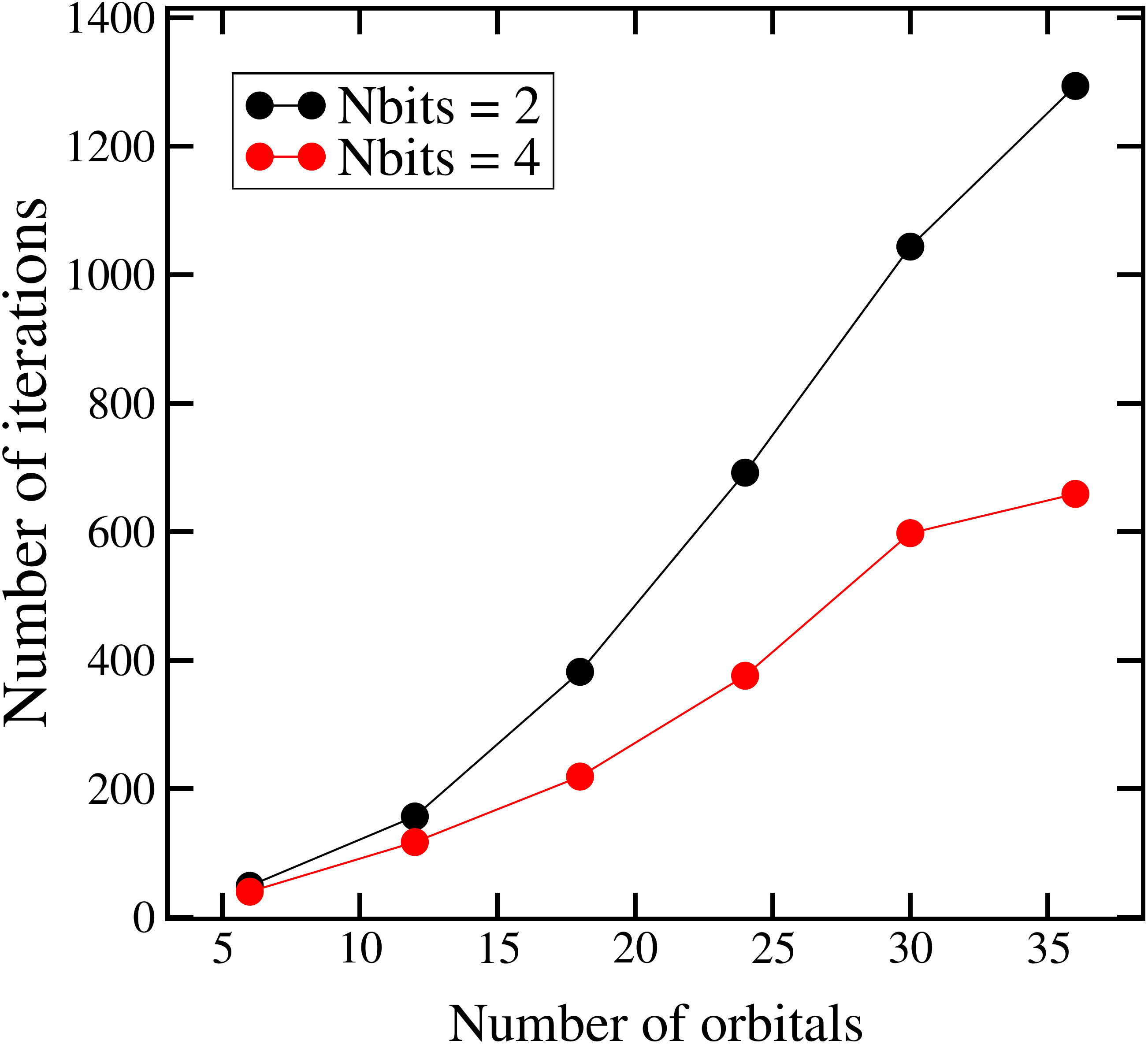} 
        \caption{Number of iterations as a function of the system size (number of orbitals). Calculations using two and four bits of precision are shown.}
        \label{its_vs_size}
    \end{figure}

    Figure \ref{its_vs_size} shows the number of iterations (total QUBO solutions) that are needed to construct the DM for two and four bits. The total number of iterations is computed by adding up the number of iterations needed to achieve the desired precision for the calculation of each eigenvector.
    From Figure \ref{err_vs_size} it does not seem that much is gained by increasing the number of bits, however, as it is shown in Figure \ref{its_vs_size}, we notice that less iterations are needed to reach convergence when using a larger number of bits. 
    
\begin{figure}[ht]
        \centering
        \includegraphics[width=7cm]{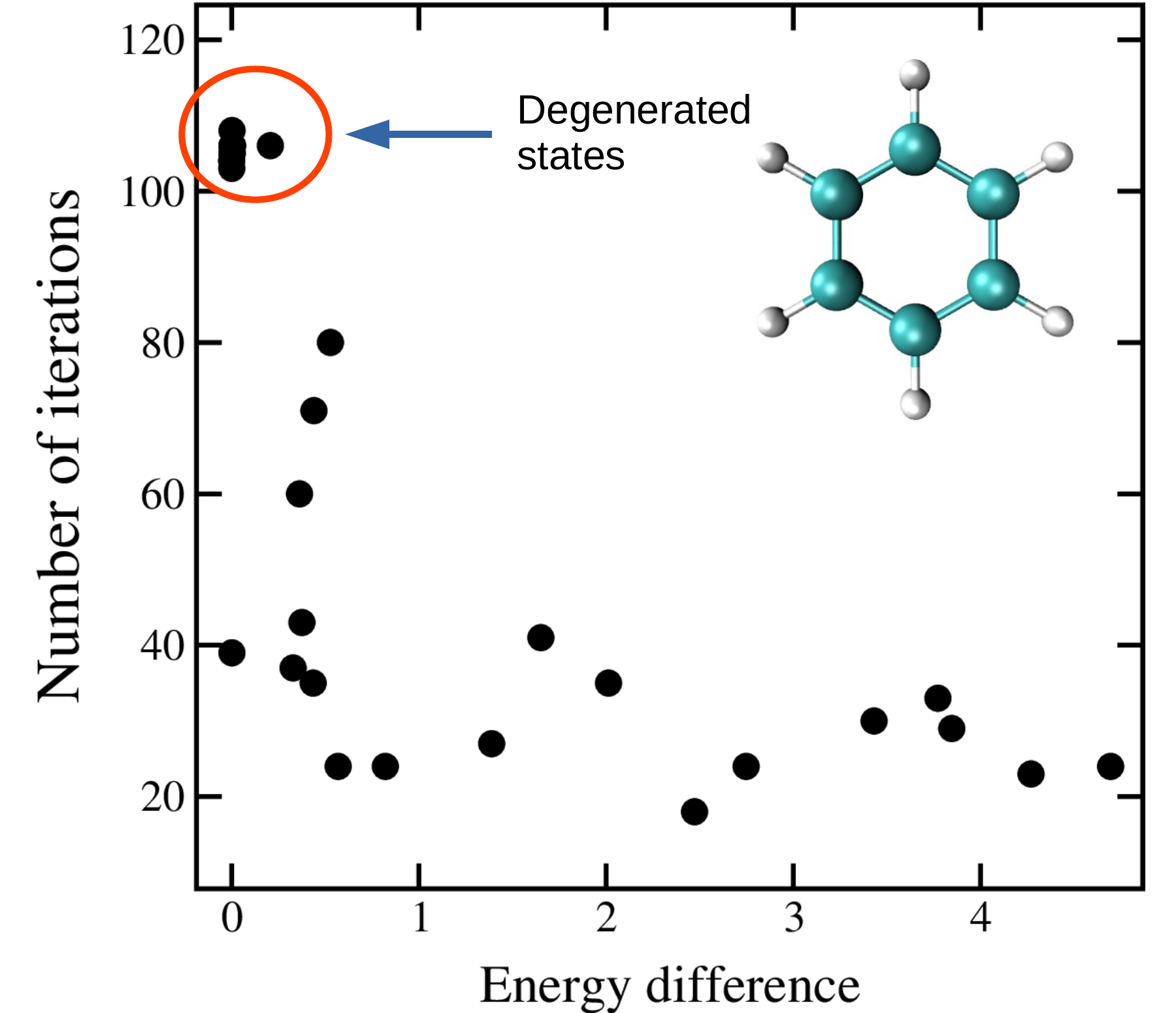}
        \caption{Number of required iterations to compute each eigenpair as a function of the minimal energy difference from other orbitals. The benzene molecule is shown in the inset with carbon and hydrogen atoms colored in cyan and white respectivelly.}
        \label{its_vs_deg}
    \end{figure}

    Finally, we computed the number of iterations needed to achieve the desired precision to compute every single state for the benzene molecule. This particular case has been selected since, due to its high symmetry (see molecular representation on the inset of figure \ref{its_vs_deg}), the electronic structure of benzene is characterized by having several groups of degenerate molecular orbitals (orbitals that are very close in energy). In general, the closer two molecular orbitals are in energy, the more the degeneracy between them. Figure \ref{its_vs_deg} shows the number of iterations for computing a single eigenvector as a function of the proximity in energy to the next eigenvector. We notice that a group of molecular orbitals that are close in energy to the next ($<$ 0.1 eV) need significantly more iterations to be computed at the desired precision.     
    This point is a significantly limiting factor since the electronic structure of several chemical systems including semiconductors and metals will have a high degree of degeneracy. For semiconductors, degeneracy will most likely happen around the band edges, while for metals it will happen around the Fermi level.
    
     \section{Quantum vs Simulated Annealing} 

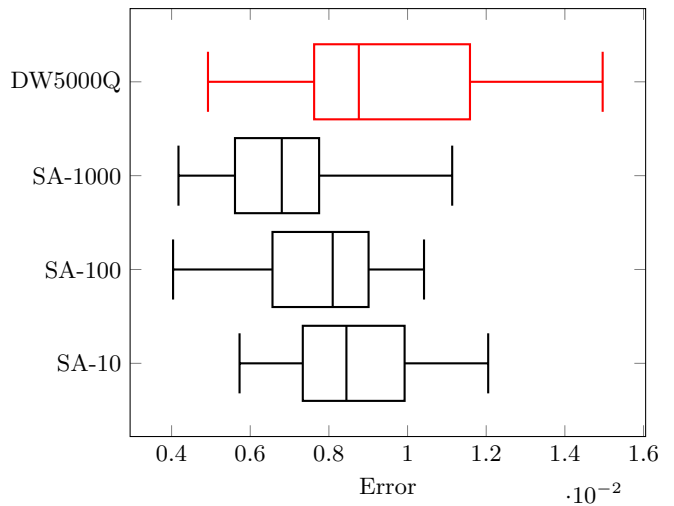
\begin{figure}[ht]  
        \centering
            \begin{tikzpicture}
  \begin{axis}
    [
    ytick={1,2,3,4,5},
    yticklabels={SA-10,SA-100,SA-1000, DW5000Q},
    xlabel={Error}
    ]
    \addplot+[color=black,thick,
        boxplot prepared={
        lower whisker= 0.005728057 , 
        lower quartile= 0.007336202 , 
        median= 0.008445314 , 
        upper quartile= 0.009924991 , 
        upper whisker= 0.012050266 
    },
    ] coordinates {};

    \addplot+[color=black,thick,
        boxplot prepared={
        lower whisker=  0.004037660, 
        lower quartile= 0.006565699 , 
        median= 0.008096661 , 
        upper quartile= 0.009008732 , 
        upper whisker=  0.010418840
    },
    ] coordinates {};
    
    \addplot+[color=black,thick,
        boxplot prepared={
        lower whisker= 0.004175210 , 
        lower quartile= 0.005611997 , 
        median= 0.006802131 , 
        upper quartile=  0.007751124 , 
        upper whisker=  0.011135036
    },
    ] coordinates {};

%     \addplot+[color=red,thick,
%         boxplot prepared={
%         lower whisker= 0.003895878 , 
%         lower quartile= 0.005404740 , 
%         median= 0.005721661 , 
%         upper quartile= 0.007855317  , 
%         upper whisker=  0.010156916
%     },
%     ] coordinates {};

        \addplot+[thick, color=red,
        boxplot prepared={
        lower whisker= 0.004925371 , 
        lower quartile= 0.007624392 , 
        median= 0.008761478 , 
        upper quartile= 0.011587146 , 
        upper whisker=  0.014961241
        },
     ] coordinates {};

  \end{axis}
\end{tikzpicture}
        \caption{Figure showing different box plots for the error commited with different methods.
        Quantum Annealer (red); and Simulating Annealing with different number of sweeps (black)}
        \label{md}
         \end{figure}

    In this section we present a comparison between the results obtained using SA with different sweeps and QA performed with the new D-Wave's Advantage 4.1 machine \cite{Pau_Farre2021-ig}. The  chain strength for QA was calculated as 30\% of the largest absolute value in the QUBO matrix \cite{chain_strength}. This helps ensure minimal chain breaks contributing to erroneous solutions. The default anneal time of 20 microseconds was used.
    %embedding: \saychris{Sue, please add description}. 
    The system DM calculation was performed over samples Hamiltonians taken from different configurations over a 100 Molecular Dynamics (MD) steps. We have extracted $H$ from 20 configurations evenly spaced 
    over the course of the MD trajectory. The DM was computed using an eigenpair precision of $10^{-5}$. From Figure \ref{md} we can see that, on average, the results using QA have a larger error. Moreover, we have observed that more iterations are needed to converge. In the case of SA a larger number of sweeps tend to converge to results with lower errors. Means for the error between SA and QA are statistically different even for small values of sweeps.   
    %
%     This shows that, albeit the algorithm to compute DM is still not competitive with tradiational methods such as regular diagonalization (Jacoby or QR methods); when using a QUBO approach QA could have an advantage over SA provided the problem can fit on the chimera graph. 
   %
    One of the mayor issues is the fact that the QUBO matrices have a high dynamic range which causes a ``resolution problem.'' When the number of bits is large there are many small QUBO entries that could have large relative error at the moment of embedding but that might have a large influence in the results. One could scale the QUBO matrix Q so that the minimun entry has a value of 1.0, but this introduces the need of having very strong chain strengths that could introduce yet another source of error. Moreover, we have observed that some chain length are too long due to the high connectivity or matrix density which also depends on the number of bits. Other types of optimization problems have also revealed no advantage using QA vs SA  \cite{Koshikawa2021-ho}.

    \section{Alternative approaches}
    
    In this section we will briefly describe some alternative approaches that could potentially overcome the issues that 
    were found by using a direct method to construct the DM using a QUBO eigensolver. Three different approaches are hereby suggested
    together with a discussion of possible issues they might have. 
    1) Using a QUBO based linear solver on the $\mathcal{O}(N)$ method recently proposed by L. Anderson \cite{Andersson2019-bz}. This algorithm shifts the  computational cost of solving $\rho$ to the solution of a linear system ($A\vect{x} = b$) using Conjugate Gradient method which can be replaced by:
 \[
 \vect{x} = \underset{\vect{x}}{\text{argmin}}\left((A\vect{x} - b)^{2} \right) 
\]
    and can be easily formulated as a QUBO problem and solved on a QA. The problem that this method could have is the fact that there might be several iterations (QUBO solves) needed to reach the desired precision of $\vect{x}$. This will add yet another inner loop into the algorithm.
    
    2) We can use Green Functions ($G(z)$) method and find $\rho$ by integrating over the complex contour (with variable $z \in \mathbb{C}$) containing the eigenspectrum of $H$. For this proposition we would need to have a method to invert a complex matrix using a QUBO formulation. Provided we can compute the inverse of a matrix using a QUBO formulation, $\rho$ would be determined as: 
    \begin{align}
     \rho = \int_C G(z) dz     
    \end{align}
    where  $G(z) = (H - zI)^{-1}$ and C is the semicircle above the real axis. Inverting a matrix $A$ can be simply done by solving $N$ linear systems of the form $A \vect{x}_i - \vect{e}_i  = 0$,
    with $\vect{e}_i$ being the i-th canonical vector, and $\vect{x}_j$ the j-th column of $A^{-1}$. This can be easily formulated as a QUBO provided both, the solution and the matrix to be inverted are real. If this is not the case, as for the Green Function method here proposed, the formulation becomes more complicated and, as far as we know, no previous work has addressed this issue.  
    
    3) Finally, we could think about solving for $\rho$ ``all at once'' from a QUBO problem where $\rho$ would be written as a long vector 
    $\vect{x}$ containing $N \times N$ entries, such that: $\vect{x}_k = \rho_{ij}$ where $i = \floor{k/N} +1$, and $j = k - N\floor{k/N}$; where $\floor{.}$ is the operator that takes the ``floor'' value of a real number. In this case, the function to be minimized will be the total energy 
    \begin{equation}
    E = \tr{\rho H} \equiv \vect{1}^{T} (H \otimes  I )\vect{x}     
    \label{superop}
    \end{equation}
    where in this case $\vect{1}_k =  \delta_{ij}$, where $i = \floor{k/N} + 1$, and $j = k - N\floor{k/N}$. Although equation \ref{superop} can be easily formulated as a QUBO problem, the energy needs to be minimized under certain constraints given by the properties of $\rho$; some of which can be extremely difficult to formulate as a QUBO. 
    Moreover, the size of the QUBO problem to be solved will scale as $N^2\times N^2$, probably opening up other bottlenecks in the algorithm even if the annealing process require no time.

    \section{Conclusion}
    
    We have demonstrated that the DM can be computed by using a QUBO based eigensolver and summing up the self outer products of the eigenvectors up to the number of occupied states.
    Although the results obtained with this direct approach are encouraging, many issues still needs to be addressed. We reported on the problem of the $\mathcal{O}(N^2)$ operation inside loops that compute the egenpairs which leads back to an $\mathcal{O}(N^3)$ scaling. 
    By experimenting with different occupation numbers, we have seen that the error rapidly increases with the occupation number regardless of the precision at which we compute each eigenvector. 
    We have also seen that the error increases with the system size regardless of the number of bits used to compute the eigenpairs. We observed that by increasing the number of bits the number of iterations to converge decreases significantly. We have identified a problem where the number of iterations to reach precision increases for degenerate states.
    Finally we showed that QA does not show any advantage when compared to SA for computing DM within this QUBO formulation and more research needs to be done to fully understand the QA vs SA comparison.
    
    We have also proposed other alternative methods to compute the DM and analyze some possible issues. 
    These methods include using linear scaling algorithms based on linear systems that could be solved using QUBO solvers;
    using Green functions obtained from a QUBO solver that then could be used to compute the DM by a close integral surrounding the eigenspectrum through the complex plane; and solving the full density matrix ``all at once'' from a QUBO problem using a vectorized form of the DM.
    Even if all these alternative methods have challenges there is still an opportunity for further improvements. 
    An in depth study of these other methods will be the subject of future work.

    \section{Acknowledgments}
    Research presented in this article was supported by the Laboratory Directed Research and De-velopment (LDRD) program of Los Alamos National Laboratory (LANL) under project number 20200056DR. This research was also supported by the U.S. Department of Energy (DOE) National Nuclear Security Administration (NNSA) Advanced Simulation and Computing (ASC) program at LANL. 
    We acknowledge the ASC  program for providing the support for accessing D-Wave's Advantage 4.1 computing resource.
    LANL is operated by Triad National Security, LLC, for the National Nuclear Security  Administration  of  U.S.  Department  of  Energy  (Contract  No.   89233218NCA000001). Assigned: LA-UR-22-20271    
    
    \section{Appendix}
    
    \subsection{Multiplication in binary representation}
    
    Multiplying $x$ and $y$ in the binary representation leads to the following operation:  
    \[
    x \times y  = q^{T}_{x} (\vect{v}^{T} \otimes \vect{v}) q_{y}   
    \]
    An example with 3-bits follows. Given we have: $x = 0.5 \equiv (0,1,0)$, and  $y = 0.25 \equiv (0,0,1)$
    \begin{align*}
        x \times y = 0.125 \\ = (0,1,0) 
        \begin{pmatrix}
            -1 & - 0.5 & -0.25\\
            -0.5 & 0.25  & 0.125\\
            - 0.25  & 0.125 &  0.0625
        \end{pmatrix}
        \begin{pmatrix}
            0  \\
            0 \\
            1 
        \end{pmatrix}
        = 0.125
    \end{align*}
    
    \subsection{Density matrix upper bound error}
    The error in $\rho$ is computed as: 
    $\Delta \rho = || \rho - \rho_{ex} || $; where $\rho_{ex}$ is the exact DM constructed using regular diagonalization methods. To ensure chemical accuracy we need to ensure that the total energy error remains below 
    1.0 kCal/mol. This means: $\Delta E < 1.0 kCal/mol $. 
    \[
    \Delta E =  \frac{\partial \tr{\rho H}}{\partial \rho} \Delta \rho +  \frac{\partial \tr{\rho H}}{\partial H} \Delta H
    \]
    Assuming no error is commited in the calculation of the Hamiltonian, we get:
    \[
    \Delta E =   \sum_{lm} \partial  \frac{  \sum_{ik}  \rho_{ik} H_{ki}    }{\partial \rho_{lm}} \Delta \rho_{lm} 
    \]
    
    \[
    \Delta E =   \sum_{lm}   H_{lm}   \Delta \rho_{lm}  \le   \max_{lm} ( H_{lm} ) \sum_{lm}  |\Delta \rho_{lm}|  
    \]
    meaning that:
    \[
    \Delta E  \le   \max_{lm} ( H_{lm} ) || \rho - \rho_{ex} ||_1 =  \max_{lm} ( H_{lm} ) \Delta  \rho 
    \]
    From the last inequality, we deduce that the maximun tolerated error in $\rho$ needs to satisfy: 
    
    \[
    \Delta E  \le   \max_{lm} ( H_{lm} ) \Delta  \rho   \le  1kCal/mol 
    \]
    then, 
    \[
    \Delta  \rho   \le  \frac{1kCal/mol}{ \max_{lm} ( H_{lm} )}
    \]

    \bibliography{mybib.bib}
    
\end{document}